\documentclass[aps,prl,twocolumn,showpacs,superscriptaddress]{revtex4}
\usepackage{graphics}
\usepackage{epsfig}

\begin{document}

\title{Network Models of Phage-Bacteria Coevolution}

\author{{\bf Martin Rosvall}, {\bf Ian B. Dodd}, {\bf Sandeep Krishna} and
{\bf Kim Sneppen}.}

\email{sneppen@nbi.dk} \affiliation{Niels Bohr Institute ,
Blegdamsvej 17, Dk 2100, Copenhagen, Denmark}
\homepage{http:/cmol.nbi.dk}
\date{\today}

\begin{abstract}
Bacteria and their bacteriophages are the most abundant, 
widespread and diverse groups 
of  biological entities on the planet. In an attempt to understand 
how the interactions between bacteria, virulent phages and temperate 
phages might affect the diversity of these groups, we developed a 
novel stochastic network model for examining the co-evolution of 
these ecologies. In our approach, nodes represent whole species or strains of 
bacteria or phages, rather than individuals, with ``speciation" and 
extinction modelled by duplication and removal of nodes. Phage-bacteria 
links represent host-parasite relationships and temperate-virulent phage 
links denote prophage-encoded resistance. 
The effect of horizontal transfer of genetic information between strains was 
also included in the dynamical rules. The 
observed networks evolved in a highly dynamic fashion but the ecosystems 
were prone to collapse (one or more entire groups going extinct). 
Diversity could be stably maintained in the model only if the probability 
of speciation was independent of the diversity. Such
an effect could be achieved in real ecosystems if the speciation rate 
is primarily set by the availability of ecological niches.
\end{abstract}
\maketitle

\noindent
{\bf Introduction}\\

Bacteria and the bacteriophages that infect them are present in huge
numbers in a wide range of natural environments, e.g.\ $\sim${}$10^6$
bacteria and $10^7$ phages per ml of seawater \cite{breibart}. 
Phages are significant factors in determining
bacterial mortality \cite{weinbauer}, and thereby have a major
influence on global recycling of nutrients and carbon in the
biosphere \cite{suttle}. 
The diversity of these populations is also staggering, with estimates of
$\sim${}$10^2$ different bacterial species and $\sim${}$10^3$ phage
genotypes per few litres of seawater \cite{breibart,weinbauer} and
at least $10^4$ phage genotypes per kg of marine sediment
\cite{breibart-b}. 
Further, because most phages only infect
very few host strains, the composition of phage strains is likely to be an
important determinant of the composition of bacterial communities.
Moreover, bacteria and phage populations are dynamic: 
they have been observed to fluctuate wildly on timescales ranging from 
weeks to months \cite{chibani}.

Natural phage populations comprise both virulent and temperate
phages. Replication of a virulent phage kills the
host bacterium (lytic life cycle), whereas a temperate phage can
replicate either lytically or by temporarily combining its genome
with that of the bacterium to form a lysogen (lysogenic life cycle).
The phage genome in a lysogen (prophage) provides its bacterial host
with immunity to lytic infection by the same strain of phage.
Deterministic predator-prey modeling \cite{levin,stewart} of
phage-bacterial ecosystems with virulent and temperate phages have shown
that these may be stable (all three classes coexisting) or unstable
(one or more classes collapsing), depending on predation and reproduction
parameters. 

In this paper we build several more coarse-grained models, consisting of
a network of nodes, representing bacterial and phage strains, and links,
representing interaction between strains, which evolves stochastically
in discrete time steps according to a set of rules. 
The nodes of the network are bacterial and phage strains, i.e., subpopulations, 
however we do not explicitly model the populations as dynamical variables.
Instead, the rules for adding or removing nodes and links use only
the structural properties of the network at that time. 
For instance, we take the extinction rate of a bacterial
strain to be a simple function of the number (and type)
of phage strains that can infect it, i.e., the number of links
pointing to it from phage strains.
In contrast, in a model where populations were modeled explicitly, the rule would
be that an extinction occurs whenever the population of a strain falls to zero.
The population, in turn, would typically be derived from a differential
equation that would depend on the number of links pointing from phages to
the given bacterial strain. In our modelling approach, we short-circuit this
step, replacing populations and their differential equations by a simple
rule based on properties like the number of links.

Our model rules incorporate various biological facts concerning phage and bacteria
interactions. For instance, we take into account the ability of temperate phages to carry
genes that make the lysogenic host resistant to infection by
virulent phages \cite{hendrix}, providing bacteria with weapons in
the co-evolutionary arms race with virulent phages. We also incorporate horizontal
transfer of genes between phages sharing the same host in the rules that determine
the evolution of new phage strains.
Since our models coarse-grain the system at the level of strains of bacteria and phage,
they are particularly suited to examine questions about the diversity of bacterial
and phage populations, rather than their sizes.\\

\begin{figure}[t]
\centering
\includegraphics[width=\columnwidth]{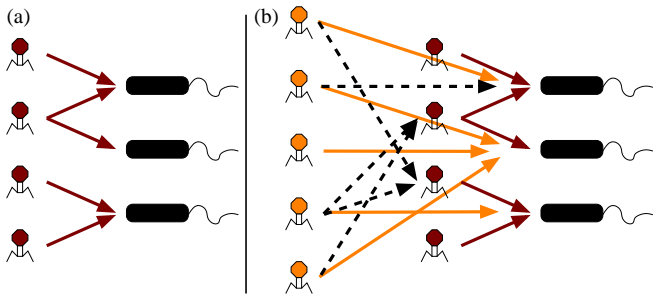}
\caption{Schematic illustration of model B (a) and model C (b). Model B consists of a
``trophic layer" of virulent phage strains (red) that can infect a layer of bacterial strains (black).
The infection possibility is indicated by the directed arrows (links).
Model C has, in addition, a layer of
temperate phage strains (orange) which can provide resistance to bacteria against the virulent
ones. This is indicated by links between temperate and virulent phage strains.
Links from either type of phage strain to bacterial strains indicate the ability 
of phage to infect the bacteria they point to.\label{fig1}}
\end{figure}

\vspace{1cm}
\noindent
{\bf Models and Results}\\

\begin{figure}[t]
\centering
\includegraphics[width=\columnwidth]{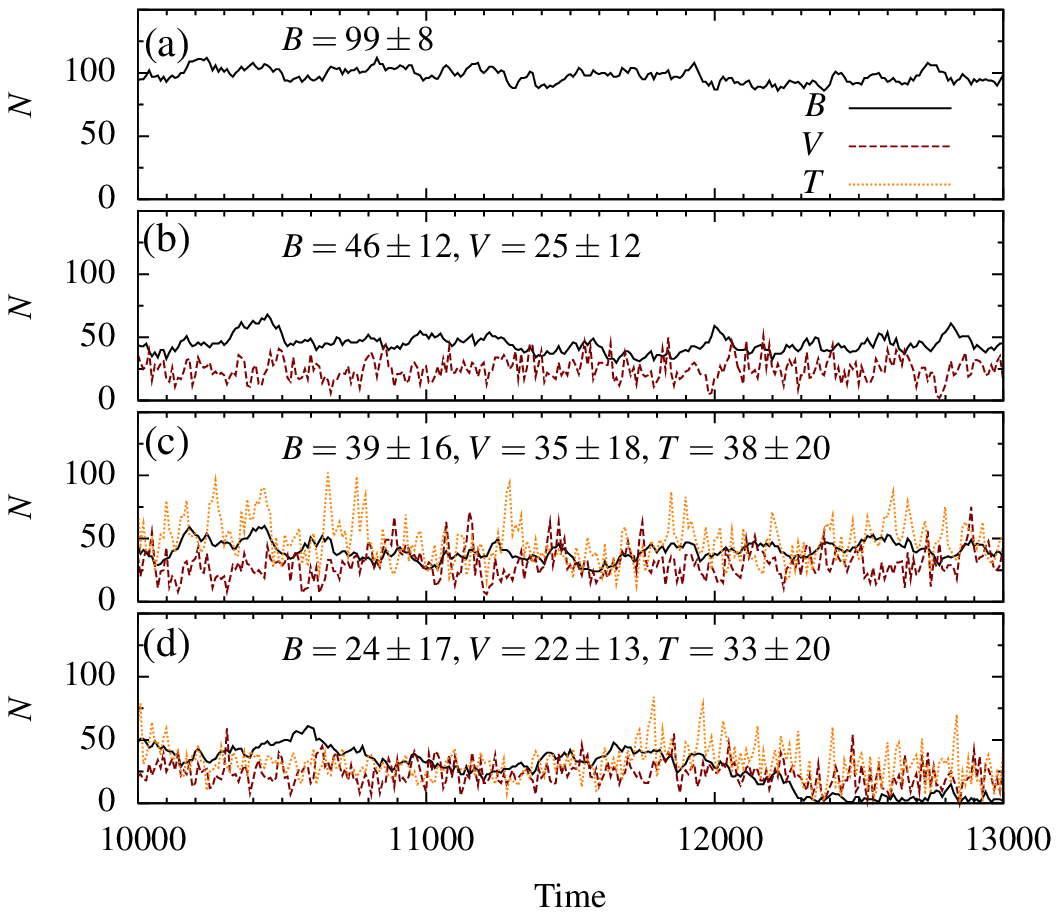}
\caption{Dynamics of model A (a), B (b), C (c), and C without the
resistance to virulent phages provided by temperate phages (d).
Parameters are $N_0=100$, $\mu=2.5$, $\beta=2.0$, and
$\sigma=0.2$. Plots show number of strains: bacterial (black), virulent (red),
temperate (orange). Accompanying numbers show average$\pm$ standard deviation
of the strain numbers for the time window shown in the figure. \label{fig2} }
\end{figure}

The model ecosystem is built on top of a ``trophic layer" of a variable
number, $N_B$, of different bacterial strains. In the absence of phages,
this number fluctuates because of the extinction of strains as well as the
creation of new ones (which we henceforth term ``speciation").
We let the speciation rate define our time step. Thus at every time, $t$,
two types of events occur to change the system: 
\begin{itemize}
\item \emph{Speciation:} We select a random strain and duplicate it, \mbox{$ N_B(t) = N_B(t-1)+1$}.
\item \emph{Extinction:} We remove each strain, $i=1,2,...,N_B(t)$, with a probability $N_B/N_{0}^2$.
\end{itemize}
The second rule implements random extinction associated with environmental loads
common to all strains, like eukaryotic predators, scarce resources and crowding.
The parameter $N_0$ represents the carrying capacity of the
ecosystem. In this simple non-interacting system 
$N_B$ fluctuates
around $N_0$, as illustrated in Fig.\ \ref{fig2}a.
Note that removing each strain with probability $1/N_0$ would produce
the same behaviour. Instead we choose $N_B/N_0^2$ to take into account the reduced extinction
rate when there are fewer strains and therefore more biomass per strain.
We denote this scenario of non-interacting bacterial strains, with no phages,
``model A".

On top of this basic system of independent bacteria
we add a self adjusting number, $N_V$, of virulent phage strains. 
This extended model B
now also contains links between phage and bacterial strains
as illustrated in Fig.\ \ref{fig1}a. Such a link represents the
ability of a phage to infect and lyse the bacteria it points to.
As before, the bacterial speciation rate defines the basic time step.
At each time, $t$, the following events occur:
\begin{itemize}
\item \emph{Bacterial speciation:} We select a random bacterial strain and
duplicate it, along with its original links, and then remove a random link,
if possible.
\item \emph{Phage speciation:} We select randomly a number of phage strains,
the number being drawn from a Poisson distribution with mean $\mu$. 
For each selected phage we create
a duplicate, copying all original links, and then
adding a link to a single bacterial strain. This bacterial strain is selected
randomly or locally (explained below) with equal weight. 
\item \emph{Extinction:} We remove each bacterial strain, $i$, 
with a probability $n/N_0^2$ (where $n$ is an effective total number of strains, explained below)
and, in addition, with a probability $\beta/N_0$ for each link from a virulent phage to that bacterial
strain.
Similarly, we remove each phage strain $j$ with a probability $\sigma/N_0$ for
each link from that phage to a bacterial strain. 
We also remove all phages that are left without any host, i.e., with zero links.
\end{itemize}

In case the number of bacterial or phage strains falls to zero, we reintroduce a single
strain with a random link to/from the other group.

The bacterial speciation rule is a simple modification from model A, the 
removal of one link representing the possibility of new strains improving
their fitness by developing resistance to existing phages.
The parameter $\mu$ spacifies the rate of phage evolution, relative to
the bacterial evolution, being the average number of new phage strains
that arise per bacterial duplication.

The addition of a link represents the possibility of a new phage strain evolving
the ability to infect a different host. 
A ``local" choice of the new host bacterial strain models horizontal transfer, between phages, of 
genes for infecting bacteria. For instance, if two phages infect the 
same bacterium then one could gain genes
from the second phage which could allow it to infect one of the latter phage's
hosts. 
We implement this by first making a list of all other phage strains that share
a common host with the phage strain just duplicated. Then we find 
all the bacterial strains having links from this set of phage strains,
but not from the duplicated phage.
Finally, we randomly choose one out of this set of bacterial strains 
and add a link to it from the duplicated phage.
For example, if the top phage (phage 1) in Fig.\ \ref{fig1}a
duplicates, the duplicate will add a link to the bacterium in the
middle because it shares the top bacterium with phage 2 which has the
middle bacterium as a host. 
Note that we make such a local choice half the time. The other half of the speciation
events are non-local, i.e., the bacterial strain is chosen randomly, 
representing evolution of new functionality in the phage, or horizontal
transfer between bacteria, which allows the phage to infect
a completely new bacterial strain.

The extinction rules are also a simple extension of model A rules.
Each link now results in a ``load", $\beta$, 
on the corresponding bacterial strain, which increases it's extinction
probability. In addition, there is an extinction rate common to all strains
given by $n/N_0^2$, with $n=\sum_{i=1}^{N_B} e^{-\beta b_i}$ replacing
$N_B$ in the extinction rule of model A ($b_i$ is the number of links from phages to
the bacterial strain $i$). Instead of taking just $N_B$ we reduce the weight of 
each bacterial strain to take into account the load from the phages that infect it.
Then, the probability that a particular bacterial strain $i$ survives is
$(1-n/N_0^2)\times (1-\beta/N_0)^{b_i}\approx (1-n/N_0^2)e^{-\beta b_i/N_0}$
when $N_0$ is large. 
The total probability for extinction of a bacterial strain due to phage load is therefore
$1-(1-n/N_0^2)e^{-\beta b_i/N_0}$.

Similarly, each link also sets a load $\sigma$ on
the phage because it should allocate genes to deal with the
strain-specific chemistry of its potential hosts. The genes may code
for proteins that change the host machinery to accommodate phage
replication, or for proteins used for the attachment or injection
of the phage genome into the host, or for proteins fighting
the countermeasures taken by the bacteria. A phage that can enter
several different strains of bacteria would need more genes, which
in turn would reduce its replication rate, here represented by a
phage extinction probability $\sigma/N_0$ per link. This is implemented
by making the net extinction probability of a phage strain
$1-e^{-\sigma v_j/N_0}$, where $v_j$ is the number of out-links it has
(this formula is similar to the bacterial extinction probability above
except the extra common extinction rate $n/N_0^2$ which does not apply to the phage strains.)

To avoid systematic errors we randomize the order in which we perform
multiple duplications as well as the order in which the strains are
selected in the extinction step.
In general we will assume that $\beta >\sigma$, reflecting a larger
load on a bacteria than on the phage infecting it. A value of the load $\beta
\sim 1$ corresponds to a situation where each virulent phage imposes a load on
the bacterial ecology which is similar to the background extinction rate
(set by $1/N_0$).

In Fig.\ \ref{fig2}b we show the dynamics of model B
with $\beta=2$, $\sigma=0.2$ and $\mu=2.5$. Comparing
with Fig.\ \ref{fig2}a the first observation is the smaller
number of bacteria, reflecting the load, $\beta$, on the bacteria imposed by
the phage. Further, the number of strains in the
ecosystem fluctuates relatively more than for model A.
This is partly expected as adding links introduces
correlations, and thus reduces the effective number of independent
variables from $\sim N_B$ to a smaller number. In addition,
the process of duplication in itself includes a positive
feedback from the number of links to itself, a feedback that is only
limited by the bound on link density set by $\sigma$.

Finally we 
introduce temperate phages, that can insert their own genome into 
their hosts' genome, forming a
lysogen. These phages kill a fraction of the
bacteria upon infection, but also leave some of them immune to
superinfection. As a consequence, they cannot drive the population of
a bacterial strain to extinction. Nevertheless they 
present a load on a
bacterial strain (i.e., affect it's extinction rate), 
in part because they often manipulate the bacteria's metabolism and
also because they increase the length of the bacterial genome and
thereby its generation time (typically one finds 0-10 prophages in
bacterial genomes \cite{casjens}). We represent this in the network
by links connecting temperate phage strains to bacterial strains (see Fig.\ \ref{fig1}b).

Another important characteristic of lysogenic phages is that they
confer immunity not only to superinfection by their own strain, but
can also provide resistance towards infection from other phages. 
We implement this 
in our model C by adding links between temperate and virulent
phages, as illustrated in Fig.\ \ref{fig1}b. Such a link implies that infection of a bacterium by that temperate
phage confers on the bacterium a resistance to infection by the virulent
phage the link points to.
(In our model we ignore temperate phages providing resistance 
to other temperate phages.)
Every link from a virulent phage to a bacterial strain is either ``strong",
if the bacteria have no link from a temperate phage that can provide resistance
to the lytic phage, or ``weak" if there does exist such a link (in the network picture,
every link from a virulent phage to a bacteria is either part of a triangle, in which case it
is weak, or not, in which case it is strong).
In contrast, every link from a temperate phage to a bacterial strain is
a weak link.
A strong link always results in a large load, $\beta$, on the bacterial strain
the link points to. A weak link, results in a weak load, $\sigma$, on the
bacterial strain. All links also result in a weak load, arbitrarily
set to $\sigma$, on the phages from which the links originate.

The final model C incorporating bacteria, virulent phages and temperate phages
is defined below.
At each time step (with a timescale set by the bacterial speciation rate)
the following events occur:
\begin{itemize}
\item
\emph{Bacterial speciation}:
We select a random bacterial strain and duplicate it by copying it
with all the originals links, and then remove one link if possible.
In choosing which link to remove we give highest priority to strong links,
then to weak links from virulent phage strains and finally to weak links
from temperate phages.
\item
\emph{Virulent phage speciation}: We choose a number of phages to duplicate, drawn
from a Poisson distribution with mean $\mu$. For each chosen phage we make
a duplicate with all the original links and then make a local or random modification
(as in model B) with equal weight. The modification is either the addition
of a link to a bacterial strain or the removal of a link from a temperate phage.
\item
\emph{Temperate phage speciation}: 
We choose a number of temperate phages to duplicate, drawn
from a Poisson distribution with mean $\mu$. For each chosen phage we make
a duplicate with all the original links and then either add a link to a 
bacterial strain or to a virulent
phage (chosen locally or randomly as in model B).
\item
\emph{Bacterial extinction}: Each bacterium $i$ is removed with 
probability $1-(1-n/N_0^2)e^{-\beta b_i/N_0}e^{-\sigma v_i/N_0}$,
where $b_i$ is the number of strong links pointing to it, $v_i$ is the number
of weak links pointing to it, and $n=\sum_{i=1}^{N_B} e^{-\beta b_i-\sigma v_i}$ is
an effective number of bacterial strains.
\item
\emph{Phage extinction}: Each phage $j$ is removed with probability 
$1-e^{-\sigma v_j/N_0}$, where $v_j$ is the number of out-links it has.
In addition, every phage without a bacterial host is removed.
\end{itemize}

This model is a straightforward extension of model B.
Speciation is assumed to occur by duplication of an existing strain
with small modifications that are likely to increase the 
fitness of that strain. Thus, for bacteria the modification is always the loss
of a link, while for virulent phages it is the gain of a new host or the
evolution of means to overcome resistance due to some temperate
phage. For a temperate phage the modification is either the gain of a new
bacterial host or the gain of genes that provide resistance (for the bacteria) to some
virulent phage. In either case, the temperate phage receives a new link, which
points to a bacterial strain or virulent phage, that is chosen locally or randomly with
equal weight. 
As in model B, a local choice means that the temperate phage gains such a
link by copying it from another phage with which it shares a common host
bacterial strain. 

We represent the
load of temperate phages on a bacterial strain by the same weak load
parameter $\sigma$ as used before. Also we use $\sigma$ to
characterize the load that the ability to infect a bacterial strain 
puts on the temperate phage
due to increased demand on the phage gene repertoire.
In short, the overall model can be described in terms of
speciation and extinction events whose rates depend on the load on
bacteria and phages. Here we simplify matters by allowing only two types
of load, ``strong" ($\beta$) and ``weak" ($\sigma$). 
Thus, the model has 3 key parameters: \\
1) $\beta/\sigma$, the ratio of strong to weak load,\\
2) $\beta$, which sets the scale of loads for links in the system,\\
3) $\mu$, the relative speciation rate of phages.\\
In addition, we have a hidden parameter in the fifty-fifty choice of local
versus random link formation in the phage speciation rules. Varying this ratio
does not affect any of our conclusions (more details are below).

\begin{figure}[t]
\centering
\includegraphics[width=\columnwidth]{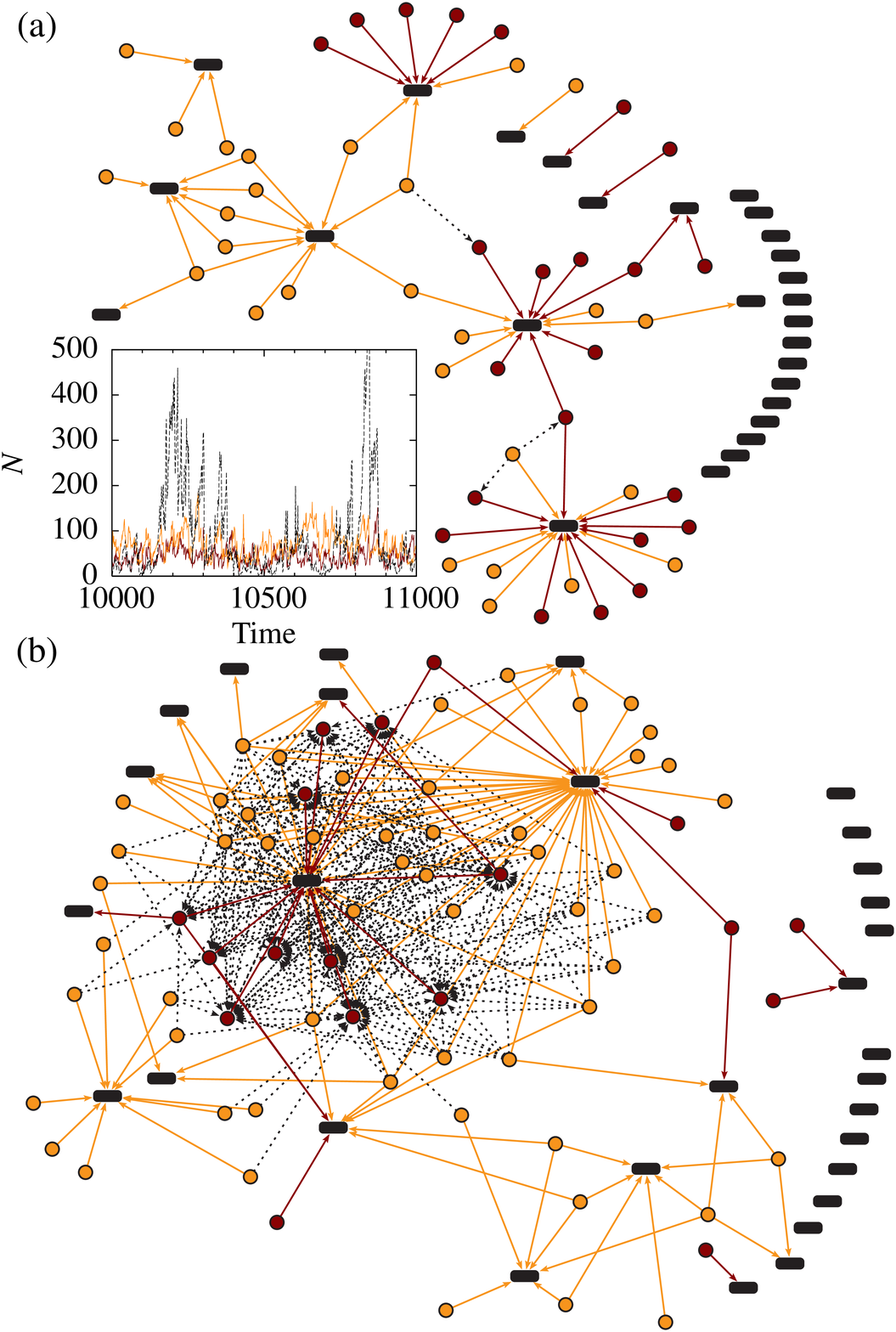}
\caption{Two examples of networks generated in the same run of model C.
Parameters are $N_0=100$, $\mu=2.5$, $\beta=2.0$ and
$\sigma=0.2$. (a) shows a network with few links and only 2 triangles
(in lower right corner). (b) shows a network with many links and
more than 100 triangles.
The inset shows the time development of the number of links in the
system for this run (color coded as in the Fig.\ \ref{fig1}). \label{fig3}}
\end{figure}

Fig.\ \ref{fig2}c illustrates the dynamics of model C. 
Comparing with Fig.\ \ref{fig2}b we see that presence
of temperate phages allows the existence of more virulent strains.
This is likely due to the lowering of the average load of virulent phages
on bacterial strains due to the
resistance provided by temperate phages. 
This conclusion is bolstered by Fig.\ \ref{fig2}d where we show the dynamics
that results when the model is modified so that temperate phages provide
no resistance (i.e., when all links from virulent phages to bacteria
are strong). This plot also shows that the resistance conferred allows
a higher number of bacterial strains to exist than when there is no resistance.

Another observation that can be made from Fig.\ \ref{fig2}c
is that the presence of temperate phages tends to increase
fluctuations. This is likely due to the intermittent increase in links from the
temperate to virulent phages, that can be seen in the inset of
Fig.\ \ref{fig3}.
The number of links from temperate to virulent phages fluctuates especially
strongly, as a result of which the network structure also varies enormously
(as evident from the network snapshots in Fig.\ \ref{fig3}). Thus, one
feature of our model is that the network structure is more variable and dynamic
than could be guessed from observing the total numbers of bacterial and phage
strains alone.
This conclusion also holds if we vary the ratio between local and random choice
of link formation in the phage speciation rule. Quantitatively, increasing the proportion
of random link formation moderately reduces the number of links from temperate to
virulent phage strains, whereas increasing the proportion of local link
formation reduces the number of bacterial strains connected to phages, leaving a larger
number of them isolated.

\begin{figure}[t]
\centering
\includegraphics[width=\columnwidth]{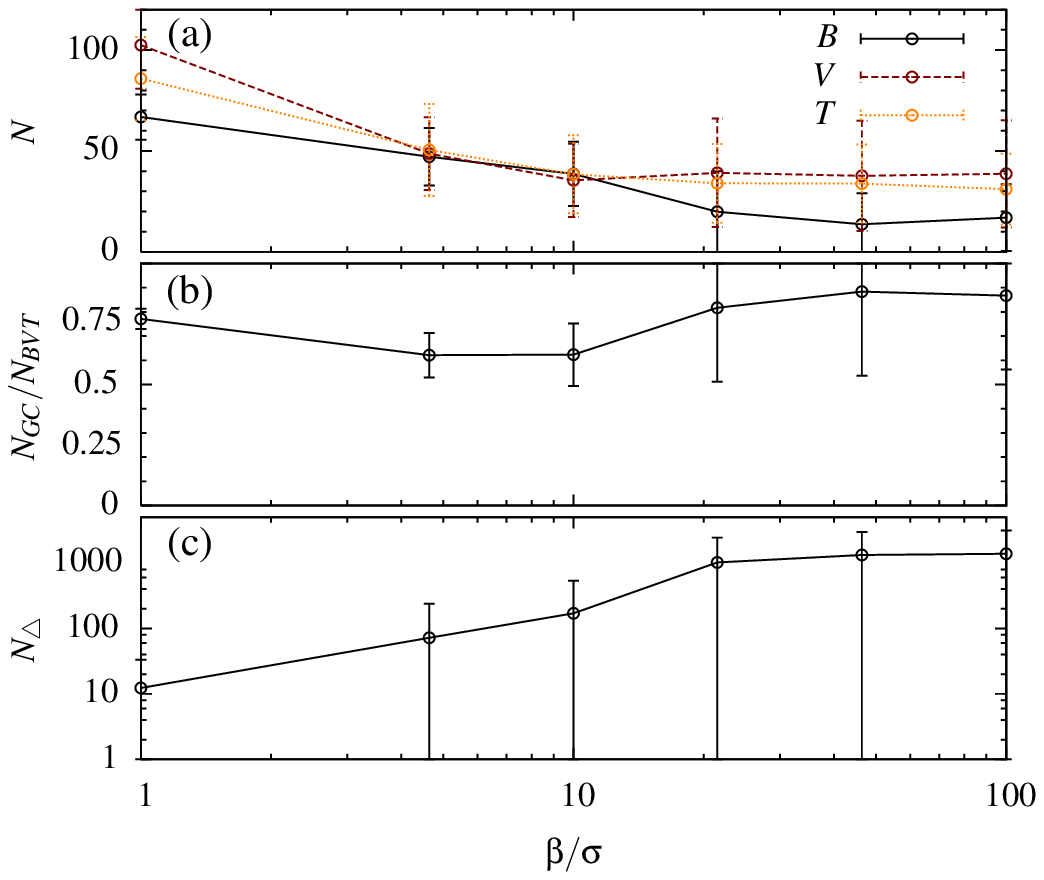}
\caption{Behaviour of model C as a function
of the ratio of strong to weak load, $\beta /\sigma$. We use a fixed
$\sigma=0.2$, $\mu=2.5$, and $N_0=100$ in the simulations and vary $\beta$. 
(a) strain numbers for bacterial, virulent and temperate groups (color coded as
in Fig.\ \ref{fig1}), (b) fractional size of the largest connected cluster in
the network, (c) number of triangles in the network. Error bars show one standard
deviation. \label{fig4}}

\end{figure}

We have examined the model against variations of the three basic
parameters, ($\beta$, $\sigma$ and $\mu$). First of all, reducing
$\beta$ and $\sigma$ while keeping $\beta/\sigma$ fixed produces an
ecology with a larger number of phages and a larger number of links
per phage. One can also increase the phage to bacteria ratio without
changing link density by assigning an especially weak load for
temperate phages on bacteria. Thus, the overall ratio of vira to
bacteria is easily re-scaled. The total size of the ecosystem, bacteria plus phages,
on the other hand, is primarily set by $N_0$.

Given fixed $\sigma$ (and fixed $\mu$) we examine, in Fig.\ \ref{fig4}, 
the behavior of the model ecology as function of the strong to weak
load ratio, $\beta/\sigma$. Fig.\ \ref{fig4}a shows that an increase in the ratio
seems to reduce the overall numbers of both bacteria and phages.
This is not surprising, since an increased ratio corresponds to an increased load $\beta$.
Temperate phages seem to fare marginally better than virulent ones 
only for intermediate values of the ratio, while bacteria do better
when the ratio is smaller. Interestingly, the 
fractional size of the largest connected
cluster in the network (shown in Fig.\ \ref{fig4}b) does
not change much though it fluctuates more for larger $\beta/\sigma$
ratios. This is a result of the increased interconnectedness at higher $\beta/\sigma$
which is revealed in the number
of triangles, shown in Fig.\ \ref{fig4}c. Note that in our model
a triangle necessarily has to be between one bacterial strain, one
virulent and one temperate phage strain, which provides resistance to
that virulent phage, i.e., the number of triangles
reflects the number of weak links between virulent phage and bacteria.
The figure suggests that resistance due to temperate phages plays a
larger role at higher $\beta/\sigma$ ratios but also that this resistance is
intermittent, with large fluctuations from time to time.
Overall, we observe a network structure that, while being usually one
large, connected cluster, is nevertheless highly dynamic as indicated
by the wildly fluctuating number of links and triangles.

\begin{figure}[t]
\centering
\includegraphics[width=\columnwidth]{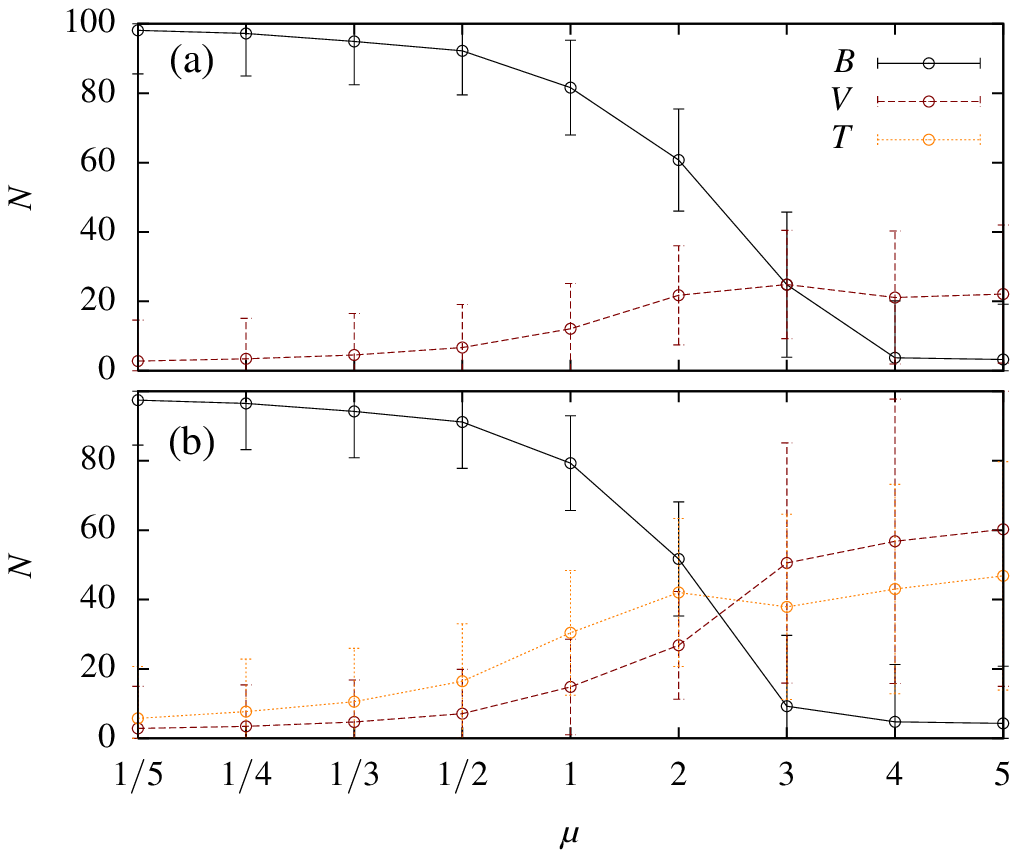}
\caption{Variation in 
phage duplication rate $\mu$. 
Other parameters are kept fixed at the same 
values as in Fig.\ \ref{fig2} and \ref{fig3}.
Plots show strain numbers for (a) model B, and (b) model C.
\label{fig5}}
\end{figure}

The last important parameter of the model is $\mu$, the speciation rate
of phages relative to that of bacterial strains. Fig.\ \ref{fig5}
shows that the state of the system is quite sensitive to this
parameter in both model B (Fig.\ \ref{fig5}a) and model C (Fig.\ \ref{fig5}b).
Not surprisingly, when $\mu$ is increased sufficiently, the number of
bacterial strains falls, while the number of phage strains increase.
What is surprising is the steepness of the fall: a threefold change
in $\mu$ (from 1 to 3) causes more than an eightfold change in bacterial
numbers for model C. 

Although the behaviour is sensitive to $\mu$, at all values the
number of virulent phage strains and the number of temperate phage strains
are comparable. This is in part due to a bias in the phage speciation rule.
Because we always add a fixed number, $\mu$, of new virulent \emph{and} new temperate
phage strains in each time step, the speciation rate per strain is not a constant.
It increases as the number of strains decreases, and this negative feedback prevents
strain numbers from becoming very small.
A more unbiased way of implementing the speciation is to make the rate per strain
constant. To investigate this scenario, we define a model D that is identical to model C in all
respects except that we modify the phage speciation rule as follows:
We choose (on average $\mu$) phages to duplicate randomly from the combined set
of temperate and virulent strains. This ensures that the probability for
selecting a phage of a given type is proportional to the
number of strains of that type. As a
consequence the duplication of phages in the larger group becomes
more likely and coexistence of the two groups becomes difficult.
This is indeed what we see from Fig.\ 6a: For standard parameters the
virulent group collapses, and only temperate strategies appear
viable. One remedy for this is to let
virulent phages speciate faster than temperate phages. This is a 
realistic assumption, both because the generation time is shorter for virulent
phages and because they often carry their own replication machinery.

Fig.\ 6 shows how virulent phage strains take over when
their speciation rate becomes substantially larger than that of the
temperate phages.
When virulent phages speciate a little over twice as fast as
temperate ones, on average both types are present in equal numbers. Interestingly, however, 
the time plot of Fig.\ 6b shows that they do not really coexist with equal diversity. Instead 
the system seems to switch back and forth between one state where
the temperate phages are very diverse while there are few virulent strains, and another where 
the virulent phages are very diverse while there are few temperate strains. 

Alternatively, we also considered a variant of model D where 
the virulent phage population
may be constantly supplemented by temperate phages that lose their
immunity region, e.g. \cite{LDB} (the opposite is probably not possible, simply
because loss of such a complicated function requires less mutations 
than gain). We find that the effect of this temperate-to-virulent switching
is very similar to allowing virulent phages to speciate faster.
That is, as the probability of mutating from being 
temperate to virulent increases, the number of virulent phage strains rises
at the cost of the other two groups.\\

\begin{figure}[t]
\centering
\includegraphics[width=\columnwidth]{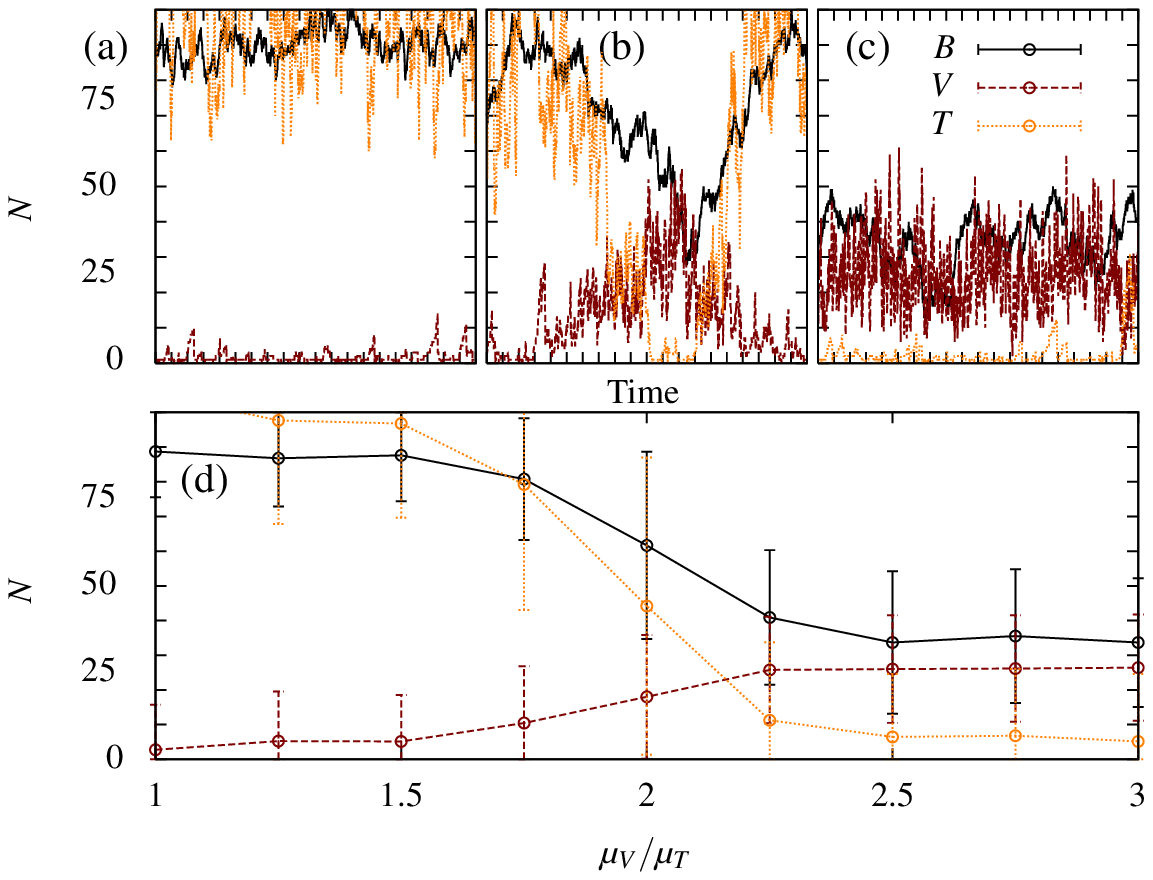}
\caption{Time course of strain numbers for (a) model D,
(b) a variant of model D where virulent strains speciate faster than temperate ones
($\mu_V=2\mu_T$), and (c) same variant with $\mu_V=3\mu_T$.
(d) shows the average strain numbers as a function of the ratio $\mu_V/\mu_T$, with
error bars showing one standard deviation.
In all plots the total phage speciation rate is fixed ($\mu_V+\mu_T=3$).
Other parameters as in Fig.\ \ref{fig2} and \ref{fig3}.
 \label{fig6}}
\end{figure}

\noindent
{\bf Discussion}\\

We have suggested a coarse-grained framework for understanding
the essential ingredients in a world governed by a co-evolutionary,
dynamical arms race between phages and their hosts. An arms race
where the elementary moves are not the fate of individual members of
the community, but rather the collapse or creation of new strains by
modification of old ones. The purpose in suggesting a mathematical model
of this kind is to remain on a level of
description that reflects our lack of knowledge of basic parameters
of infection probability and replication rates in real world
ecologies. 

Phage-bacterial ecologies are in fact quite extensive on our 
planet and govern a major fraction of the known
biomass: There are about $5 \times 10^{30}$ procaryotes on the
planet, and viral infection is the most common way in which bacteria
die, especially in the ocean.
However, exceedingly little is known about this very interesting 
part of life on our planet. 
The one basic fact that a model of phage-bacterial ecologies can attempt
to reproduce is the high diversity and coexistence of temperate and virulent phage
strains.

In practice, ecological models, whether based on population dynamics or a more
network-like approach, have great difficulty in producing viable and diverse
ecosystems where many different species and strategies coexist.
For instance, in population dynamics models that use Lotka-Volterra or replicator equations
a handful of species can coexist for a short while but are soon destroyed by
parasites\cite{JMSmith,NHB,HS}. The main reason for this is the exponential growth of self-replicating
populations that results when replication rates are proportional to the
population size. This typically results in a ``winner-take-all" situation where
the population of a slightly faster growing species can completely repress the
other populations. Only when limits are applied (sometimes artificially) on the exponential
growth can species coexist \cite{JKlargeext}.

One of the reasons we chose a network-like approach to modelling phage-bacterial ecologies,
rather than a population dynamics approach, was to try and circumvent this problem of
exponential growth and coexistence. However, our work shows that even in these kinds of models
coexistence of species is not easy to achieve. We have tried several variants
of the basic models, all within the same framework described above, and found that the
phage speciation rule was a major determinant of the viability of coexistence of
temperate and virulent phages. Model D, which makes the straightforward assumption
that the speciation rate per strain is constant, does not in fact exhibit
robust coexistence of virulent and temperate phage strains. 
There, at best, at some
carefully fine-tuned ratio of virulent to temperate speciation rates, we find 
one group present at high diversity and the other at low diversity with 
a constant switching back and forth between these states.
We also tried other variants of the rule, 
for instance where the speciation rate
of phage strains is a fixed multiple of the bacterial 
speciation rate, but this scenario 
turned out to be even worse for co-existence. 
That is, even in the absence of temperate phages either all groups
go extinct or only bacterial strains survive.

We only succeeded in achieving coexistence in the model C, where the speciation
rate of each phage group is independent of its diversity.
This ``solution" parallels the solution to the problem
of exponential growth in population dynamics models. In model D, the speciation
rate per strain is a constant, therefore the rate of increase of
strains (ignoring extinction for the moment) is proportional to the number
of strains. Thus, the number of strains would grow exponentially resulting
in a similar winner-take-all situation, now at the strain level.
In model C, however, by making the speciation rate independent of strain
number, the growth is no longer exponential and we see coexistence of a
large number of strains. 
Thus, one ``prediction" resulting from our modelling is that there may
be some mechanism at work that keeps the speciation rate independent
of the number of strains. We speculate that this might happen if 
speciation involves the discovery of new ecological niches by randomly mutated individuals.
If the number of such new ecological niches
is small then it could be what limits the speciation rate, rather than the population size.
In that case the speciation rate would be independent of strain numbers.

Since model C was the one case where we did find robust coexistence,
we mainly focus on how different phage groups influence
each other in that model. The main result of this
analysis was that:

1) Temperate phage strains in fact
appear to help maintain a higher diversity of virulent strains
by providing a ``refuge" for a few strains of
bacteria to escape to, preventing them from being completely
destroyed by virulent phages,

2) The ecosystem is highly dynamic, especially its
network structure. In particular, the number of
links between temperate and virulent phages (and hence triangles)
show large intermittent fluctuations. In other words, periods 
where bacterial strains are largely protected from virulent phages
alternate with periods where there is little resistance and
most virulent attacks present huge
extinction risks for the bacterial strains. Thus, the stabilization provided
by temperate phages is sometimes very important, and at other times
nearly without consequence.

We emphasize that these two results only hold for model C, where
each phage group produces a given fixed number of new strains at each
timestep of the model. \\

\noindent
{\bf Outlook}\\

The difficulty of finding a model where phage types coexist 
indicates that we nevertheless miss some important insight into how such ecosystems 
actually work on this very basic information-exchange level.
One intriguing possibility is that mutation mechanisms
and speciation rates could themselves change and adjust as the network evolves.
For instance, one could imagine that if viral phage strains
were allowed to evolve their speciation rate, 
they would die out both in clusters where they had too low a rate ($\mu<1/2$ in Fig.\ 5b)
and in clusters where they had too high a rate by forcing their
hosts, the bacterial strains, to collapse ($\mu>3$ in Fig.\ 5b).
The result might be the self-organization of speciation rates to values that allow
coexistence of all groups.

More realistic scenarios could also consider interactions between
temperate phage species. For example prophages can confer
resistance not only to virulent phages but also to temperate phages.
Another feature is phage-independent genetic transfer between bacteria such as
mediated by bacterial conjugation. We have loosely tried to take this into account
by the random allocation of new links from time to time. However, this
could be implemented more carefully in a non-random manner.

Overall, we have presented a flexible framework for modelling 
phage-bacterial interactions. By working at strain level, ignoring
detailed population dynamics, these models are particularly suited
for producing questions related to the diversity 
of different groups in the ecosystem.\\

\noindent
{\bf Acknowledgements}\\
This work was supported by the Danish National Research Foundation.

\end{document}